\documentclass{article} 
\usepackage{iclr2024_conference,times}

\usepackage{amsmath,amsfonts,bm}

\def\eqref#1{equation~\ref{#1}}

\def\1{\bm{1}}

\DeclareMathAlphabet{\mathsfit}{\encodingdefault}{\sfdefault}{m}{sl}
\SetMathAlphabet{\mathsfit}{bold}{\encodingdefault}{\sfdefault}{bx}{n}

\usepackage{hyperref}
\usepackage{url}
\usepackage{subcaption}
\usepackage{amsmath}
\usepackage{array}
\usepackage{graphicx}
\usepackage{xcolor}
\usepackage{geometry}
\usepackage{array}
\usepackage{colortbl}
\usepackage{multirow}
\usepackage{caption}

\definecolor{lightyellow}{rgb}{1.0, 1.0, 0.8}
\definecolor{lightred}{rgb}{1.0, 0.8, 0.8}

\title{Towards Robust Audio Deepfake Detection: A Evolving Benchmark for Continual Learning}

\author{Xiaohui Zhang \\
Institute of Automation \\ 
Chinese Academy of Sciences \\
\texttt{21120320@bjtu.edu.cn} \\
\And
Jiangyan Yi \\
Institute of Automation \\ 
Chinese Academy of Sciences \\
\texttt{jiangyan.yi@nlpr.ia.ac.cn} \\
\AND
Jianhua Tao \\
Department of Automation \\
Tsinghua University \\
\texttt{jhtao@tsinghua.edu.cn}
}

\newcommand{\ours}{EVDA}

\iclrfinalcopy 
\begin{document}

\maketitle
\begin{abstract}
    The rise of advanced large language models such as GPT-4, GPT-4o, and the Claude family has made fake audio detection increasingly challenging. Traditional fine-tuning methods struggle to keep pace with the evolving landscape of synthetic speech, necessitating continual learning approaches that can adapt to new audio while retaining the ability to detect older types. Continual learning, which acts as an effective tool for detecting newly emerged deepfake audio while maintaining performance on older types, lacks a well-constructed and user-friendly evaluation framework. To address this gap, we introduce \ours, a benchmark for evaluating continual learning methods in deepfake audio detection. \ours\ includes classic datasets from the Anti-Spoofing Voice series, Chinese fake audio detection series.
It supports various continual learning techniques, such as Elastic Weight Consolidation (EWC), Learning without Forgetting (LwF), and recent continual learning methods for audio deepfake detection like Regularized Adaptive Weight Modification (RAWM) and Radian Weight Modification (RWM). Additionally, \ours\ facilitates the development of robust algorithms by providing an open interface for integrating new continual learning methods. Our code will be available at \url{https://github.com/Cecile-hi/Evolving-FAD-CL-Benchmark.git}.
\end{abstract}
\section{Introduction}
Fake audio detection has become increasingly crucial in recent years due to the rapid development of deep learning models, particularly large language models such as GPT-4 \citep{DBLP:journals/corr/gpt4}, GPT-4o, and Claude family \citep{anthropic2024claude}. These models have significantly blurred the line between real and generated speech, making it challenging for human ears to distinguish between them. With the accessibility of tools like GPT-4o to the general public, generating synthetic audio has become effortless, presenting significant challenges for societal security and safety.
\par
The continuous development of large language models like GPT-4, GPT-4o, and others has led to the emergence of new techniques for generating synthetic speech. This evolution has made it increasingly difficult for humans to discern between generated and authentic speech. The availability of GPT-4o to free users has further democratized the creation of fake audio, presenting both convenience and significant challenges for various domains such as security.
\par
Continual learning, particularly in the context of fake audio detection, has emerged as an effective approach to address the evolving landscape of synthetic speech generation. By continuously collecting data from newly emerging techniques, including those generated by large language models, continual learning enables models to adapt to new types of synthetic speech while retaining the ability to detect previously encountered ones.

For example, traditional model training and fine-tuning techniques may result in the forgetting of older models' characteristics once the model learns to distinguish newer models like GPT-4o. In contrast, continual learning ensures that the model retains the ability to discern older models like GPT-4 while learning to identify newer ones.
\par
To address the challenges posed by evolving large models and speech generation techniques, we propose a continually updated benchmark for \textbf{EV}oloving synthetic and \textbf{D}eepfake \textbf{A}udio detection using various continual learning methods, named \textbf{\ours}. Our benchmark includes classic datasets from the Anti-Spoofing Voice series challenges \citep{wu2015asvspoof, todisco2019asvspoof, asv2021} for detection tasks, as well as Chinese fake audio detection datasets such as Audio Deepfake Detection series (ADD 2022, ADD 2023)\citep{yi2022add, DBLP:conf/dada/YiTFYWWZZZRXZGW23}, among others and newly emerging deepfake audio generated by recent large language models, such GPT-4 and GPT-4o.

Additionally, we plan to incorporate fake audio generated by models like GPT-4, GPT-4o, Claude, and future large models as new detection tasks. On the continual learning front, we support classic methods like Elastic Weight Consolidation (EWC)\citep{kirkpatrick2017overcoming}, Learning without Forgetting (LwF)\citep{li2017learning}, as well as recent methods like Regularized Adaptive Weight Modification (RAWM)\citep{DBLP:conf/icml/ZhangYTWZ23}, Radian Weight Modification (RWM)\citep{DBLP:conf/aaai/ZhangYWZZ024}. Furthermore, we provide an open interface to allow researchers to integrate their own continual learning methods into the benchmark, facilitating the development of more effective and robust algorithms to tackle the continuous evolution of fake audio generated by large models.

The primary contribution of this paper is \ours, a benchmark designed to evaluate continual learning methods in the context of evolving deepfake audio detection. \ours\ encompasses both classic and newly emerging datasets, and supports various continual learning algorithms. We emphasize the importance of continual learning as a solution to address the dynamic nature of these challenges.

\section{Related Work}

\subsection{Audio Deepfake Detection}
The rise of synthetic audio, often referred to as deepfake audio, has led to significant advancements in detection techniques aimed at distinguishing between genuine and manipulated audio content. Early methods primarily focused on spectral analysis and traditional machine learning approaches, leveraging features such as Mel-frequency cepstral coefficients (MFCCs) and Gaussian Mixture Models (GMMs) to detect anomalies indicative of synthetic audio \citep{wu2015asvspoof}. With the evolution of deep learning, more sophisticated models, including convolutional neural networks (CNNs) and recurrent neural networks (RNNs), have been developed to capture subtle temporal and spectral differences between real and fake audio. Recent advancements have introduced hybrid models that combine deep learning with signal processing techniques to improve detection accuracy in challenging environments, such as noisy or low-quality recordings \citep{todisco2019asvspoof, asv2021}. Additionally, cross-lingual and cross-dataset generalization has become a focal point, with new datasets and benchmarks being established to test models against a wider range of synthetic audio scenarios \citep{yi2022add, muller2022does}.

\subsection{Continual Learning Methods}
Continual learning, also known as lifelong learning, addresses the challenge of training models on sequential tasks while retaining knowledge from previous tasks. In the context of deepfake audio detection, this capability is crucial as new types of attacks and spoofing techniques emerge. Traditional models suffer from catastrophic forgetting, where learning new tasks leads to the erosion of previously acquired knowledge. To mitigate this, methods such as Elastic Weight Consolidation (EWC) have been proposed, which constrain the updates of important parameters to preserve knowledge from earlier tasks \citep{kirkpatrick2017overcoming}. Other techniques, such as experience replay \citep{DBLP:conf/nips/Replay} and knowledge distillation \citep{li2017learning}, have also been employed to maintain performance across tasks by rehearsing previous data or transferring knowledge from a stable teacher model to a student model during training. Recent advancements in continual learning have focused on adapting these methods to complex neural architectures, such as transformers and deep CNNs, making them more robust in dynamic environments where the nature of the data and tasks evolves over time \citep{DBLP:conf/cvpr/iCaRL}. In the context of deepfake detection, these continual learning approaches \citep{DBLP:conf/icml/ZhangYTWZ23, DBLP:conf/aaai/ZhangYWZZ024} are pivotal in ensuring that models can adapt to new challenges without (or miner) losing the ability to detect previously known spoofing techniques.

\section{Proposed Method}
To address the growing complexity and diversity of audio deepfake detection, we propose a comprehensive benchmark designed to rigorously evaluate the performance of anti-spoofing models across a variety of challenging scenarios. Existing audio deepfake detection methods often focus on limited conditions or datasets, failing to capture the full range of real-world variations that models may encounter \citep{ma2021continual, DBLP:conf/icml/ZhangYTWZ23, DBLP:conf/aaai/ZhangYWZZ024}. Our benchmark aims to fill this gap by incorporating eight distinct tasks, each carefully selected to represent different data sources, languages, and acoustic conditions. This diversity allows for a thorough assessment of model robustness, adaptability, and cross-lingual capabilities, pushing the boundaries of current audio spoofing detection methods. The benchmark is structured to challenge models with varying degrees of difficulty, ensuring that only the most advanced and resilient models can succeed across all tasks.
\subsection{Data Architecture}
Our benchmark architecture consists of eight tasks (Task 1 to Task 8), each representing different datasets and conditions for detecting synthetic audio. These tasks vary in data sources, language, acoustic conditions, and keep the number of training and evaluation samples in same. The data architecture of \ours\ benchmark are listed as Table \ref{tab: arch-bench}, which is designed to comprehensively evaluate the performance of audio spoofing and anti-spoofing models across a diverse set of conditions and datasets. Below, we detail each experiment's setup and objectives:
\par
\begin{table}[t]
\centering
\caption{Data architecture of our proposed benchmark}
\label{tab: arch-bench}
\resizebox{\linewidth}{!}{
    \begin{tabular}{|c|c|c|c|c|c|c|}
    \hline
    \textbf{ID} & \textbf{Source} & \textbf{Have Vocoder ID} & \textbf{Language} & \textbf{Acoustic Condition} & \textbf{Train} & \textbf{Test} \\
    \hline
    \textbf{Task 1} & FMFCC-A & F & Chinese & Human & 2000 & 5000 \\
    \hline
    \textbf{Task 2} & In-the-Wild & F & \cellcolor{gray!25}English & \cellcolor{gray!25}Real world & 2000 & 5000 \\
    \hline
    \textbf{Task 3} & ADD 2022 & F & \cellcolor{gray!25}Chinese & \cellcolor{gray!25}Low-quality/Partially fake & 2000 & 5000 \\
    \hline
    \textbf{Task 4} & ASVspoof2015 & T & \cellcolor{gray!25}English & \cellcolor{gray!25}No significant noise & 2000 & 5000 \\
    \hline
    \textbf{Task 5} & ASVspoof2019LA & T & English & \cellcolor{gray!25}Logical access & 2000 & 5000 \\
    \hline
    \textbf{Task 6} & ASVspoof2021LA & T & English & Logical access & 2000 & 5000 \\
    \hline
    \textbf{Task 7} & FoR & F & English & \cellcolor{gray!25}Human & 2000 & 5000 \\
    \hline
    \textbf{Task 8} & HAD & F & \cellcolor{gray!25}Chinese & \cellcolor{gray!25}Human/Partially fake & 2000 & 5000 \\
    \hline
    \end{tabular}}
\end{table}

 \textbf{Task 1: FMFCC-A (Chinese)}\citep{DBLP:conf/iwdw/FMFCC-A} - Utilizes 2000 training and 5000 evaluation samples of Chinese audio. This experiment is critical for models to adapt to low-quality, partially fake audio environments, increasing the difficulty of the spoofing detection task.
    
 \textbf{Task 2: In-the-Wild (English)}\citep{muller2022does} - Uses 2000 training and 5000 evaluation samples in real-world audio scenarios, testing the models' effectiveness in dynamically changing and challenging acoustic environments.

 \textbf{Task 3: ADD 2022 (Chinese)}\citep{yi2022add} - Incorporates 2000 training and 5000 evaluation samples of Chinese audio. This experiment focuses on low-quality, partially fake settings, presenting challenges in detecting spoofed audio.

 \textbf{Task 4: ASVspoof2015 (English)}\citep{wu2015asvspoof} - Utilizes 2000 training and 5000 evaluation samples from the ASVspoof2015 dataset, focusing on English language audio in clean conditions without significant channel or background noise effects. Vocoder ID is available, aiding in more precise synthetic speech detection.

 \textbf{Task 5: ASVspoof2019 - Logical Access (LA) (English)}\citep{todisco2019asvspoof} - Incorporates 2000 training and 5000 evaluation samples focusing on logical access scenarios to test models against sophisticated spoofing attacks in English. Vocoder ID is again available, providing critical data for assessing synthetic signal traits.

 \textbf{Task 6: ASVspoof2021 - Logical Access (LA) (English)}\citep{asv2021} - Continues the logical access examination with the latest dataset iteration, using English audio samples. The presence of vocoder ID helps in identifying new spoofing techniques that have emerged since the 2019 dataset.

 \textbf{Task 7: FoR (English)}\citep{DBLP:conf/for/ReimaoT19} - Uses 2000 training and 5000 evaluation samples of purely human-generated audio in English, testing the robustness of models against non-synthetic audio scenarios.

 \textbf{Task 8: HAD (Chinese)}\citep{DBLP:conf/interspeech/HAD} - Utilizes 2000 training and 5000 evaluation samples of Chinese audio, which includes a mix of human and partially fake audio, presenting challenges in cross-lingual and authenticity discernment.



\subsection{Arrangement of the tasks}
We have arranged and highlighted the table data based on the distinct characteristics of each experiment on Table \ref{tab: arch-bench}, emphasizing the differences in language and acoustic conditions. This organization allows for a clearer comparison of how various datasets and conditions affect the performance of deepfake audio detection models. By highlighting the changes, we can better understand the specific challenges each dataset introduces and how the models need to adapt to these variations.
The key distinctions in the experiments are reflected in the changes to both \textbf{language} and \textbf{acoustic conditions} as follows:

- \textbf{Task 1 to Task 2:} The language changes from Chinese to English, with the acoustic condition shifting from "Human" to "Real world." This transition highlights the model's need to adapt from a controlled environment to more natural, varied conditions.

- \textbf{Task 2 to Task 3:} The language switches back to Chinese, while the acoustic condition changes to "Low-quality/Partially fake," indicating a mix of real and synthetic audio in more challenging quality conditions.

- \textbf{Task 3 to Task 4:} The language changes to English, with the acoustic condition becoming "No significant noise," focusing on clean audio without significant interference.

- \textbf{Task 4 to Task 5:} The language remains English, but the focus shifts to "Logical access" scenarios, introducing sophisticated spoofing attacks.

- \textbf{Task 5 to Task 6:} The language and acoustic condition remain consistent (English, "Logical access"), but the dataset iteration updates to the latest version with newly emerged spoofing technologies.

- \textbf{Task 6 to Task 7:} The language stays the same (English), while the acoustic condition returns to "Human," emphasizing purely human-generated audio.

- \textbf{Task 7 to Task 8:} The language changes back to Chinese, with the acoustic condition now being "Human/Partially fake," involving a mix of human and synthetic audio.

These highlighted changes are crucial for understanding the adaptability and robustness of deepfake audio detection models across different datasets, languages, and acoustic conditions.
The benchmark is designed not only to assess the current state-of-the-art in spoofing detection but also to push the development of more robust, adaptable, and cross-lingual models capable of handling the complexities of real-world audio spoofing scenarios.

\subsection{Data Selection Methodology}
For each task within our benchmark, we provided two strategies for data selection:

 \textbf{Random Selection:} We randomly select a fixed number of authentic and spoofed audio samples from the original datasets or collected data. Initially, we split the data into two groups: Real and Fake. From the "Real" group, we randomly select 1000 real samples, and from the "Fake" group, we select 1000 fake samples corresponding to various generation methods. These samples are combined to form the training set. For the evaluation set, we follow the same procedure but select only 2500 real and 2500 fake samples. This approach ensures a balanced representation of real and fake samples in both the training and evaluation sets. In current benchmark, this method has been used for data selection.

 \textbf{Most Informative Selection:} In this method, we aim to select the most informative samples from the original datasets. First, we train \( n \) discriminative models for spoof detection using external datasets, ensuring these models achieve extremely low loss on their respective training sets. These well-trained models act as experts during the sample selection process. Each sample from the original dataset is evaluated by all experts, and the samples are scored based on the entropy of the experts' predictions using the formula:
    \begin{equation}
        \text{Entropy} = - (p_1 \log p_1 + p_2 \log p_2)
    \end{equation}
where
    \begin{equation}
        p_1 = \frac{N_{\text{real}}}{N_{\text{real}} + N_{\text{fake}}}, \quad p_2 = \frac{N_{\text{fake}}}{N_{\text{real}} + N_{\text{fake}}}    
    \end{equation}
Here, \( N_{\text{real}} \) and \( N_{\text{fake}} \) represent the number of models classifying the sample as real or fake, respectively. According to information theory \citep{shannon1948mathematical}, samples with the highest entropy values contain the most information, indicating uncertainty and hence, difficulty in classification. We list the entropy of each sample and rank them from highest to lowest. The top 2500 real and 2500 fake samples are selected to construct the evaluation set, ensuring it includes the most challenging and informative samples. The next 1000 real and 1000 fake samples are selected for the training set, which benefits from the inclusion of diverse and informative samples.
\subsection{Continuous Learning Methods in proposed Benchmark}
The benchmark evaluates various continual learning methods to enhance the adaptability of audio spoofing detection systems. These approaches are summarized as follows:

\textbf{Replay} \citep{DBLP:conf/nips/Replay} mitigates catastrophic forgetting by replaying a subset of previously seen samples during training, helping to maintain performance on earlier tasks.

\textbf{Finetuning} adapts a pre-trained model to new tasks by continuing training with new data, effectively leveraging existing models to incorporate new knowledge.

\textbf{EWC} \citep{kirkpatrick2017overcoming} adds a regularization penalty to the loss function based on parameter importance, preserving critical knowledge from previous tasks.

\textbf{GDumb} \citep{DBLP:conf/eccv/GDumb} combines finetuning with periodic retraining on a data subset, balancing new and old information for large-scale learning.

\textbf{CWRStar} \citep{DBLP:journals/nn/CWRStar} employs rehearsal and interleaving tasks using a fixed memory, improving retention of past knowledge while learning new tasks.

\textbf{SI} \citep{DBLP:conf/icml/SI} introduces dynamic regularization based on the importance of model parameters, balancing the retention of old knowledge with the acquisition of new information.

\textbf{OWM} \citep{zeng2019continual} adjusts weight updates to remain orthogonal to those associated with previous tasks, reducing forgetting by preserving relevant features.

\textbf{RAWM} \citep{DBLP:conf/icml/ZhangYTWZ23} adapts weight updates based on the ratio of genuine to fake utterances, introducing regularization to retain knowledge across different acoustic conditions.

\textbf{RWM} \citep{DBLP:conf/aaai/ZhangYWZZ024} uses radian-based weight modification to minimize disruption to previously learned features, enhancing stability and adaptability in continual learning.

In summary, our benchmark provides a comprehensive evaluation of these methods, aiming to develop robust models that can effectively handle evolving audio spoofing attacks with high reliability and accuracy.

\section{Result}
\subsection{Compared with multiply continual learning methods}

\begin{table}[ht]
\centering
\begin{tabular}{c|c|c|c|c|c|c|c|c|c}
\hline
\textbf{Methods} & \textbf{Task1} & \textbf{Task2} & \textbf{Task3} & \textbf{Task4} & \textbf{Task5} & \textbf{Task6} & \textbf{Task7} & \textbf{Task8} & \textbf{Avg} \\ 
\hline
Replay & \cellcolor{lightred}2.88 & \cellcolor{lightred}7.60 & \cellcolor{lightred}2.56 & \cellcolor{lightred}1.40 & \cellcolor{lightred}2.80 & \cellcolor{lightred}8.84 & \cellcolor{lightred}6.76 & \cellcolor{lightred}4.12 & \cellcolor{lightred}4.62 \\ 
Finetuning & 8.28 & 10.04 & 5.36 & \cellcolor{lightyellow}1.52 & \cellcolor{lightyellow}4.08 & 18.24 & 10.44 & \cellcolor{lightyellow}4.44 & 7.80 \\ 
EWC & \cellcolor{lightyellow}6.64 & 10.16 & 5.52 & 1.72 & 4.20 & 16.92 & \cellcolor{lightyellow}10.08 & 4.56 & \cellcolor{lightyellow}7.475 \\ 
GDumbFinetune & \cellcolor{lightred}2.88 & \cellcolor{lightyellow}8.28 & \cellcolor{lightyellow}4.92 & 3.64 & 5.88 & 19.28 & 13.16 & 10.44 & 8.56 \\ 
CWRStar & 28.2 & 36.72 & 22.48 & 29.8 & 46.4 & 49.36 & 47.04 & 34.12 & 36.77 \\ 
SI & 8.52 & 10.2 & 7.48 & 1.68 & 5.40 & 17.12 & 13.92 & 4.60 & 8.615 \\ 
OWM & 6.72 & 28.52 & 32.88 & 14.32 & 19.84 & 40.8 & 26.08 & 21.88 & 23.88 \\ 
RAWM & 9.24 & 13.6 & 5.38 & 1.68 & 4.36 & \cellcolor{lightyellow}16.4 & 12.44 & 6.00 & 8.638 \\ 
RWM & 9.64 & 14.76 & 8.52 & 2.04 & 4.64 & 17.6 & 11.88 & 5.96 & 9.38 \\
\hline
\end{tabular}
\caption{EER (\%) on each task for different continual learning methods. The cells with the lowest EER in each column are highlighted in red, while the cells with the second-lowest EER are highlighted in yellow. Lower EER values indicate better performance.}
\label{tab:continual_learning}
\end{table}
Table \ref{tab:continual_learning} summarizes the performance of various continual learning methods across eight tasks, evaluated by Equal Error Rate (EER) \citep{wu2015asvspoof}. Compared with all continual learning methods, replay consistently shows most competitive performance across all tasks, demonstrating its overall effectiveness. Finetuning and EWC exhibit relatively similar performances. However, EWC achieves the better performance in old and average tasks, indicating its overcoming forgetting performance in old tasks. GDumbFinetune performs well in certain tasks, notably achieving the lowest EER in Task 1 along with Replay. However, its overall performance is less consistent, with a higher average EER. CWRStar consistently shows higher EER values across all tasks, indicating it is less effective compared to other methods, with the highest average EER (36.77).SI and OWM demonstrate mixed results. SI achieves low EERs in some tasks, but not as consistently as Replay or EWC. OWM shows the highest EER in several tasks, particularly in Task2 and Task3. RAWM and RWM provide relatively competitive performances but fall behind Replay and EWC in several tasks. RAWM has a notable second-lowest EER in Task 6 compared to the best.
Overall, Replay and EWC are the top-performing methods, with Replay showing the lowest average EER. These results suggest that Replay and EWC are effective in minimizing error rates in continual learning scenarios for audio deepfake detection, while other methods like CWRStar and OWM may need further refinement for better performance. 
\subsection{Continual learning on each task}
We evaluate our benchmark using the EER metric on 8 tasks using the conventional Finetuning and EWC continual learning method. All methods and tasks are trained using 5 linear layers model with 128-hidden dimension. The EER results are illustrated as Table \ref{tab:combined-evaluation}:
\begin{table}[ht]
\centering
\resizebox{0.85\linewidth}{!}{
\begin{tabular}{c|c|c|c|c|c|c|c|c|c}
\hline
\multicolumn{2}{c}{\textbf{Train after}} & \textbf{Task1} & \textbf{Task2} & \textbf{Task3} & \textbf{Task4} & \textbf{Task5} & \textbf{Task6} & \textbf{Task7} & \textbf{Task8} \\
\hline
\multirow{2}{*}{Task1} & FT & 5.44 & 32.12 & 28.52 & 33.80 & 42.12 & 39.20 & 45.00 & 22.32 \\
\cline{2-10}
& EWC & 5.44 & 32.12 & 28.52 & 33.80 & 42.12 & 39.20 & 45.00 & 22.32 \\
\hline
\multirow{2}{*}{Task2} & FT & 10.20 & 6.32 & 25.84 & 7.72 & 10.64 & 30.60 & 21.92 & 12.28 \\
\cline{2-10}
& EWC & 9.48 & 6.20 & 26.72 & 8.96 & 12.24 & 29.92 & 23.00 & 12.24 \\
\hline
\multirow{2}{*}{Task3} & FT & 11.48 & 15.04 & 1.48 & 19.16 & 44.40 & 34.92 & 46.28 & 19.72 \\
\cline{2-10}
& EWC & 12.92 & 15.36 & 1.64 & 20.40 & 49.80 & 36.84 & 42.76 & 20.16 \\
\hline
\multirow{2}{*}{Task4} & FT & 16.56 & 13.00 & 4.12 & 1.80 & 6.32 & 25.76 & 17.36 & 9.92 \\
\cline{2-10}
& EWC & 17.96 & 12.80 & 4.80 & 1.72 & 6.16 & 27.32 & 16.76 & 9.92 \\
\hline
\multirow{2}{*}{Task5} & FT & 18.56 & 10.28 & 9.28 & 1.76 & 2.64 & 23.60 & 12.08 & 11.08 \\
\cline{2-10}
& EWC & 21.04 & 9.96 & 10.60 & 1.68 & 2.36 & 23.88 & 12.16 & 10.96 \\
\hline
\multirow{2}{*}{Task6} & FT & 10.04 & 12.36 & 5.64 & 2.40 & 6.96 & 9.96 & 14.12 & 11.16 \\
\cline{2-10}
& EWC & 10.12 & 11.52 & 6.20 & 2.16 & 6.52 & 10.04 & 13.84 & 11.52 \\
\hline
\multirow{2}{*}{Task7} & FT & 10.84 & 12.72 & 8.20 & 1.72 & 2.64 & 12.44 & 6.64 & 12.40 \\
\cline{2-10}
& EWC & 11.08 & 11.96 & 9.72 & 2.04 & 2.52 & 12.84 & 6.28 & 12.64 \\
\hline
\multirow{2}{*}{Task8} & FT & 8.28 & 10.04 & 5.36 & 1.52 & 4.08 & 18.24 & 10.44 & 4.44 \\
\cline{2-10}
& EWC & 6.64 & 10.16 & 5.52 & 1.72 & 4.20 & 16.92 & 10.08 & 4.56 \\
\hline
\end{tabular}
}
\caption{The EER(\%) evaluated on all tasks after each training using conventional finetuning (FT) and EWC.}
\label{tab:combined-evaluation}
\end{table}

The results presented in Table \ref{tab:combined-evaluation} compare the EER (\%) across all tasks after each training session using two methods: conventional finetuning (FT) and Elastic Weight Consolidation (EWC). From the table, it is evident that both methods exhibit similar performance when applied to the first task, as expected, since no prior knowledge needs to be preserved. However, as training progresses on subsequent tasks, noticeable differences emerge between the two methods. For Task 2, EWC shows slightly improved performance over FT in both Task 1 and Task 2, with reductions in EER ranging from minor differences. These improvements suggest that EWC effectively mitigates catastrophic forgetting by preserving knowledge from earlier tasks, allowing for more stable learning. However, there are cases, like Task 3, where EWC does not offer significant improvements over FT, indicating that its effectiveness can be task-dependent. Overall, while both methods show varying levels of performance across different tasks, EWC generally provides a modest improvement over FT on old tasks. This highlights advantage in continual learning where the model needs to maintain performance across multiple tasks without revisiting them.

    
    
    
    
    
\section{Discussion}

In this work, we have proposed a continual learning benchmark, which is the first as best as we know, for audio deepfake detection.
One of the key advantages of our work is the thorough evaluation of continual learning methods in a challenging domain-shift condition. The varied and complex nature of audio data, combined with the evolving techniques used in generating deepfakes, makes it a particularly demanding area for continual learning. Our results provide valuable insights into the current capabilities of these methods and their applicability to real-world problems.

However, our study also has certain limitations. Firstly, the scope of continual learning methods explored in this work is limited. Although we have focused on well-established methods like FT and EWC, there are several other approaches that could offer additional benefits, such as replay-based methods, regularization techniques, and meta-learnin approaches. Furthermore, the benchmark tasks utilized in this study, while representative, may not fully capture the diversity of audio deepfakes encountered in practice.

In future work, we plan to expand our investigation to include a broader range of continual learning methods. This will not only provide a more complete understanding of the strengths and weaknesses of different approaches but also enable the proposal of novel methods tailored specifically to the challenges of audio deepfake detection. Additionally, we intend to incorporate newly emerging audio deepfake techniques, such as those powered by advanced generative models like GPT-4 and its successors, into our evaluation framework. By doing so, we aim to ensure that our models remain robust and effective in detecting the latest deepfake techniques, thereby maintaining their relevance in the face of rapidly advancing technologies.
\section{Conclusion}
In conclusion, the rapid evolution of audio generation models poses significant challenges for fake audio detection. However, by adopting continual learning approaches and establishing a comprehensive benchmark, we can effectively address these challenges. Our proposed benchmark provides a standardized platform for evaluating detection methods against evolving synthetic audio techniques, facilitating the development of more robust and effective detection algorithms.

\bibliography{iclr2024_conference}

\begin{thebibliography}{23}
\providecommand{\natexlab}[1]{#1}
\providecommand{\url}[1]{\texttt{#1}}
\expandafter\ifx\csname urlstyle\endcsname\relax
  \providecommand{\doi}[1]{doi: #1}\else
  \providecommand{\doi}{doi: \begingroup \urlstyle{rm}\Url}\fi

\bibitem[Anthropic(2024)]{anthropic2024claude}
AI~Anthropic.
\newblock The claude 3 model family: Opus, sonnet, haiku.
\newblock \emph{Claude-3 Model Card}, 2024.

\bibitem[Kirkpatrick et~al.(2017)Kirkpatrick, Pascanu, Rabinowitz, Veness, Desjardins, Rusu, Milan, Quan, Ramalho, Grabska-Barwinska, et~al.]{kirkpatrick2017overcoming}
James Kirkpatrick, Razvan Pascanu, Neil Rabinowitz, Joel Veness, Guillaume Desjardins, Andrei~A Rusu, Kieran Milan, John Quan, Tiago Ramalho, Agnieszka Grabska-Barwinska, et~al.
\newblock Overcoming catastrophic forgetting in neural networks.
\newblock \emph{Proceedings of the national academy of sciences}, 114\penalty0 (13):\penalty0 3521--3526, 2017.

\bibitem[Li \& Hoiem(2017)Li and Hoiem]{li2017learning}
Zhizhong Li and Derek Hoiem.
\newblock Learning without forgetting.
\newblock \emph{IEEE transactions on pattern analysis and machine intelligence}, 40\penalty0 (12):\penalty0 2935--2947, 2017.

\bibitem[Liu et~al.(2023)Liu, Wang, Sahidullah, Patino, Delgado, Kinnunen, Todisco, Yamagishi, Evans, Nautsch, and Lee]{asv2021}
Xuechen Liu, Xin Wang, Md~Sahidullah, Jose Patino, Héctor Delgado, Tomi Kinnunen, Massimiliano Todisco, Junichi Yamagishi, Nicholas Evans, Andreas Nautsch, and Kong~Aik Lee.
\newblock Asvspoof 2021: Towards spoofed and deepfake speech detection in the wild.
\newblock \emph{IEEE/ACM Transactions on Audio, Speech, and Language Processing}, 31:\penalty0 2507--2522, 2023.
\newblock \doi{10.1109/TASLP.2023.3285283}.

\bibitem[Ma et~al.(2021)Ma, Yi, Tao, Bai, Tian, and Wang]{ma2021continual}
Haoxin Ma, Jiangyan Yi, Jianhua Tao, Ye~Bai, Zhengkun Tian, and Chenglong Wang.
\newblock Continual learning for fake audio detection.
\newblock \emph{arXiv preprint arXiv:2104.07286}, 2021.

\bibitem[Maltoni \& Lomonaco(2019)Maltoni and Lomonaco]{DBLP:journals/nn/CWRStar}
Davide Maltoni and Vincenzo Lomonaco.
\newblock Continuous learning in single-incremental-task scenarios.
\newblock \emph{Neural Networks}, 116:\penalty0 56--73, 2019.
\newblock \doi{10.1016/J.NEUNET.2019.03.010}.
\newblock URL \url{https://doi.org/10.1016/j.neunet.2019.03.010}.

\bibitem[M{\"u}ller et~al.(2022)M{\"u}ller, Czempin, Dieckmann, Froghyar, and B{\"o}ttinger]{muller2022does}
Nicolas~M M{\"u}ller, Pavel Czempin, Franziska Dieckmann, Adam Froghyar, and Konstantin B{\"o}ttinger.
\newblock Does audio deepfake detection generalize?
\newblock \emph{arXiv preprint arXiv:2203.16263}, 2022.

\bibitem[OpenAI(2023)]{DBLP:journals/corr/gpt4}
OpenAI.
\newblock {GPT-4} technical report.
\newblock \emph{CoRR}, abs/2303.08774, 2023.
\newblock \doi{10.48550/ARXIV.2303.08774}.
\newblock URL \url{https://doi.org/10.48550/arXiv.2303.08774}.

\bibitem[Prabhu et~al.(2020)Prabhu, Torr, and Dokania]{DBLP:conf/eccv/GDumb}
Ameya Prabhu, Philip H.~S. Torr, and Puneet~K. Dokania.
\newblock Gdumb: {A} simple approach that questions our progress in continual learning.
\newblock In Andrea Vedaldi, Horst Bischof, Thomas Brox, and Jan{-}Michael Frahm (eds.), \emph{Computer Vision - {ECCV} 2020 - 16th European Conference, Glasgow, UK, August 23-28, 2020, Proceedings, Part {II}}, volume 12347 of \emph{Lecture Notes in Computer Science}, pp.\  524--540. Springer, 2020.
\newblock \doi{10.1007/978-3-030-58536-5\_31}.
\newblock URL \url{https://doi.org/10.1007/978-3-030-58536-5\_31}.

\bibitem[Rebuffi et~al.(2017)Rebuffi, Kolesnikov, Sperl, and Lampert]{DBLP:conf/cvpr/iCaRL}
Sylvestre{-}Alvise Rebuffi, Alexander Kolesnikov, Georg Sperl, and Christoph~H. Lampert.
\newblock icarl: Incremental classifier and representation learning.
\newblock In \emph{2017 {IEEE} Conference on Computer Vision and Pattern Recognition, {CVPR} 2017, Honolulu, HI, USA, July 21-26, 2017}, pp.\  5533--5542. {IEEE} Computer Society, 2017.
\newblock \doi{10.1109/CVPR.2017.587}.
\newblock URL \url{https://doi.org/10.1109/CVPR.2017.587}.

\bibitem[Reimao \& Tzerpos(2019)Reimao and Tzerpos]{DBLP:conf/for/ReimaoT19}
Ricardo Reimao and Vassilios Tzerpos.
\newblock For: {A} dataset for synthetic speech detection.
\newblock In Corneliu Burileanu and Horia{-}Nicolai Teodorescu (eds.), \emph{2019 International Conference on Speech Technology and Human-Computer Dialogue, SpeD 2019, Timisoara, Romania, October 10-12, 2019}, pp.\  1--10. {IEEE}, 2019.
\newblock \doi{10.1109/SPED.2019.8906599}.
\newblock URL \url{https://doi.org/10.1109/SPED.2019.8906599}.

\bibitem[Rolnick et~al.(2019)Rolnick, Ahuja, Schwarz, Lillicrap, and Wayne]{DBLP:conf/nips/Replay}
David Rolnick, Arun Ahuja, Jonathan Schwarz, Timothy~P. Lillicrap, and Gregory Wayne.
\newblock Experience replay for continual learning.
\newblock In Hanna~M. Wallach, Hugo Larochelle, Alina Beygelzimer, Florence d'Alch{\'{e}}{-}Buc, Emily~B. Fox, and Roman Garnett (eds.), \emph{Advances in Neural Information Processing Systems 32: Annual Conference on Neural Information Processing Systems 2019, NeurIPS 2019, December 8-14, 2019, Vancouver, BC, Canada}, pp.\  348--358, 2019.
\newblock URL \url{https://proceedings.neurips.cc/paper/2019/hash/fa7cdfad1a5aaf8370ebeda47a1ff1c3-Abstract.html}.

\bibitem[Shannon(1948)]{shannon1948mathematical}
Claude~Elwood Shannon.
\newblock A mathematical theory of communication.
\newblock \emph{The Bell system technical journal}, 27\penalty0 (3):\penalty0 379--423, 1948.

\bibitem[Todisco et~al.(2019)Todisco, Wang, Vestman, Sahidullah, Delgado, Nautsch, Yamagishi, Evans, Kinnunen, and Lee]{todisco2019asvspoof}
Massimiliano Todisco, Xin Wang, Ville Vestman, Md~Sahidullah, H{\'e}ctor Delgado, Andreas Nautsch, Junichi Yamagishi, Nicholas Evans, Tomi Kinnunen, and Kong~Aik Lee.
\newblock Asvspoof 2019: Future horizons in spoofed and fake audio detection.
\newblock \emph{arXiv preprint arXiv:1904.05441}, 2019.

\bibitem[Wu et~al.(2015)Wu, Kinnunen, Evans, Yamagishi, Hanil{\c{c}}i, Sahidullah, and Sizov]{wu2015asvspoof}
Zhizheng Wu, Tomi Kinnunen, Nicholas Evans, Junichi Yamagishi, Cemal Hanil{\c{c}}i, Md~Sahidullah, and Aleksandr Sizov.
\newblock Asvspoof 2015: the first automatic speaker verification spoofing and countermeasures challenge.
\newblock In \emph{Sixteenth annual conference of the international speech communication association}, 2015.

\bibitem[Yi et~al.(2021)Yi, Bai, Tao, Ma, Tian, Wang, Wang, and Fu]{DBLP:conf/interspeech/HAD}
Jiangyan Yi, Ye~Bai, Jianhua Tao, Haoxin Ma, Zhengkun Tian, Chenglong Wang, Tao Wang, and Ruibo Fu.
\newblock Half-truth: {A} partially fake audio detection dataset.
\newblock In Hynek Hermansky, Honza Cernock{\'{y}}, Luk{\'{a}}s Burget, Lori Lamel, Odette Scharenborg, and Petr Motl{\'{\i}}cek (eds.), \emph{Interspeech 2021, 22nd Annual Conference of the International Speech Communication Association, Brno, Czechia, 30 August - 3 September 2021}, pp.\  1654--1658. {ISCA}, 2021.
\newblock \doi{10.21437/INTERSPEECH.2021-930}.
\newblock URL \url{https://doi.org/10.21437/Interspeech.2021-930}.

\bibitem[Yi et~al.(2022)Yi, Fu, Tao, Nie, Ma, Wang, Wang, Tian, Bai, Fan, et~al.]{yi2022add}
Jiangyan Yi, Ruibo Fu, Jianhua Tao, Shuai Nie, Haoxin Ma, Chenglong Wang, Tao Wang, Zhengkun Tian, Ye~Bai, Cunhang Fan, et~al.
\newblock Add 2022: the first audio deep synthesis detection challenge.
\newblock In \emph{ICASSP 2022-2022 IEEE International Conference on Acoustics, Speech and Signal Processing (ICASSP)}, pp.\  9216--9220. IEEE, 2022.

\bibitem[Yi et~al.(2023)Yi, Tao, Fu, Yan, Wang, Wang, Zhang, Zhang, Zhao, Ren, Xu, Zhou, Gu, Wen, Liang, Lian, Nie, and Li]{DBLP:conf/dada/YiTFYWWZZZRXZGW23}
Jiangyan Yi, Jianhua Tao, Ruibo Fu, Xinrui Yan, Chenglong Wang, Tao Wang, Chu~Yuan Zhang, Xiaohui Zhang, Yan Zhao, Yong Ren, Le~Xu, Junzuo Zhou, Hao Gu, Zhengqi Wen, Shan Liang, Zheng Lian, Shuai Nie, and Haizhou Li.
\newblock {ADD} 2023: the second audio deepfake detection challenge.
\newblock In Jianhua Tao, Haizhou Li, Jiangyan Yi, and Cunhang Fan (eds.), \emph{Proceedings of the Workshop on Deepfake Audio Detection and Analysis co-located with 32th International Joint Conference on Artificial Intelligence {(IJCAI} 2023), Macao, China, August 19, 2023}, volume 3597 of \emph{{CEUR} Workshop Proceedings}, pp.\  125--130. CEUR-WS.org, 2023.
\newblock URL \url{https://ceur-ws.org/Vol-3597/paper21.pdf}.

\bibitem[Zeng et~al.(2019)Zeng, Chen, Cui, and Yu]{zeng2019continual}
Guanxiong Zeng, Yang Chen, Bo~Cui, and Shan Yu.
\newblock Continual learning of context-dependent processing in neural networks.
\newblock \emph{Nature Machine Intelligence}, 1\penalty0 (8):\penalty0 364--372, 2019.

\bibitem[Zenke et~al.(2017)Zenke, Poole, and Ganguli]{DBLP:conf/icml/SI}
Friedemann Zenke, Ben Poole, and Surya Ganguli.
\newblock Continual learning through synaptic intelligence.
\newblock In Doina Precup and Yee~Whye Teh (eds.), \emph{Proceedings of the 34th International Conference on Machine Learning, {ICML} 2017, Sydney, NSW, Australia, 6-11 August 2017}, volume~70 of \emph{Proceedings of Machine Learning Research}, pp.\  3987--3995. {PMLR}, 2017.
\newblock URL \url{http://proceedings.mlr.press/v70/zenke17a.html}.

\bibitem[Zhang et~al.(2023)Zhang, Yi, Tao, Wang, and Zhang]{DBLP:conf/icml/ZhangYTWZ23}
Xiaohui Zhang, Jiangyan Yi, Jianhua Tao, Chenglong Wang, and Chu~Yuan Zhang.
\newblock Do you remember? overcoming catastrophic forgetting for fake audio detection.
\newblock In Andreas Krause, Emma Brunskill, Kyunghyun Cho, Barbara Engelhardt, Sivan Sabato, and Jonathan Scarlett (eds.), \emph{International Conference on Machine Learning, {ICML} 2023, 23-29 July 2023, Honolulu, Hawaii, {USA}}, volume 202 of \emph{Proceedings of Machine Learning Research}, pp.\  41819--41831. {PMLR}, 2023.
\newblock URL \url{https://proceedings.mlr.press/v202/zhang23au.html}.

\bibitem[Zhang et~al.(2024)Zhang, Yi, Wang, Zhang, Zeng, and Tao]{DBLP:conf/aaai/ZhangYWZZ024}
Xiaohui Zhang, Jiangyan Yi, Chenglong Wang, Chu~Yuan Zhang, Siding Zeng, and Jianhua Tao.
\newblock What to remember: Self-adaptive continual learning for audio deepfake detection.
\newblock In Michael~J. Wooldridge, Jennifer~G. Dy, and Sriraam Natarajan (eds.), \emph{Thirty-Eighth {AAAI} Conference on Artificial Intelligence, {AAAI} 2024, Thirty-Sixth Conference on Innovative Applications of Artificial Intelligence, {IAAI} 2024, Fourteenth Symposium on Educational Advances in Artificial Intelligence, {EAAI} 2014, February 20-27, 2024, Vancouver, Canada}, pp.\  19569--19577. {AAAI} Press, 2024.
\newblock \doi{10.1609/AAAI.V38I17.29929}.
\newblock URL \url{https://doi.org/10.1609/aaai.v38i17.29929}.

\bibitem[Zhang et~al.(2021)Zhang, Gu, Yi, and Zhao]{DBLP:conf/iwdw/FMFCC-A}
Zhenyu Zhang, Yewei Gu, Xiaowei Yi, and Xianfeng Zhao.
\newblock {FMFCC-A:} {A} challenging mandarin dataset for synthetic speech detection.
\newblock In Xianfeng Zhao, Alessandro Piva, and Pedro~Comesa{\~{n}}a Alfaro (eds.), \emph{Digital Forensics and Watermarking - 20th International Workshop, {IWDW} 2021, Beijing, China, November 20-22, 2021, Revised Selected Papers}, volume 13180 of \emph{Lecture Notes in Computer Science}, pp.\  117--131. Springer, 2021.
\newblock \doi{10.1007/978-3-030-95398-0\_9}.
\newblock URL \url{https://doi.org/10.1007/978-3-030-95398-0\_9}.

\end{thebibliography}
\bibliographystyle{iclr2024_conference}

\end{document}